\batchmode
\makeatletter
\def\input@path{{/home/josip/Documents/Momentum_eq/}}
\makeatother
\documentclass[10pt,british]{elsarticle}
\usepackage[T1]{fontenc}
\usepackage[latin9]{inputenc}
\usepackage{geometry}
\usepackage{mathtools}
\usepackage{amsmath}
\usepackage{cancel}
\usepackage{setspace}
\usepackage{babel}
\begin{document}

\begin{frontmatter}{}

\title{Strong formulations of the generalised Navier-Stokes momentum equation}

\author[fesb]{Josip Baši\'{c}\corref{cor1}}

\ead{jobasic@fesb.hr}

\author[fesb]{Martina Baši\'{c}}

\author[fesb]{Branko Blagojevi\'{c}}

\cortext[cor1]{Corresponding author}

\address[fesb]{Faculty of Electrical Engineering, Mechanical Engineering and Naval
Architecture, University of Split, R. Boskovica 32, 21000 Split, Croatia}
\begin{abstract}
In this paper, the strong formulation of the generalised Navier-Stokes
momentum equation is investigated. Specifically, the formulation of
shear--stress divergence is investigated, due to its effect on the
performance and accuracy of computational methods. It is found that
the term may be expressed in two different ways. While the first formulation
is commonly used, the alternative derivation is found to be potentially
more convenient for direct numerical manipulation. The alternative
formulation relocates a part of strain information under the variable-coefficient
Laplacian operator, thus making future computational schemes potentially
simpler with larger time-step sizes.
\end{abstract}

\end{frontmatter}{}

\section{Introduction}

The motion of incompressible viscous fluids with variable viscosity
is governed by generalised Navier--Stokes (GNS) equations. Existence
of weak solution for the GNS equations is well researched \citep{Wu2012,DeOliveira2013,Liu2021},
and variational formulations are attractive since they contain only
first-order derivatives \citep{Pacheco2021} if the viscosity is solved
by projection \citep{Deteix2019}. On the other hand, the research
on solution methods for strong formulations of GNS equations for variable--viscosity
flows is sparse. Peng \emph{et al}. \citep{Peng2021} presented how
the strong form of generalised Navier-Stokes equations in Lagrangian
context may be used to simulate granular materials, and Baši\'{c}
\emph{et al}. \citep{Basic2021} extended the proposed method to simulate
any viscoplastic material. During the research, it was found that
the crucial point that defines the performance and accuracy of a strong-form
GNS computational method is the discretisation of the shear--stress
divergence. This paper introduces two perspectives on preparing the
GNS momentum equation for its direct discretisation.

The mass conservation (continuity or incompressibility) equation for
an incompressible fluid defines that the divergence of velocity is
always zero:

\begin{equation}
\nabla\cdot\boldsymbol{u}=0,\label{eq:continuity}
\end{equation}
where $\boldsymbol{u}=f\left(t,\boldsymbol{x}\right)$ is the velocity
vector field, dependent on time $t$ and location $\boldsymbol{x}$.
The GNS momentum equation is written as:
\begin{equation}
\frac{\mathrm{D}\left(\rho\boldsymbol{u}\right)}{\mathrm{D}t}=-\nabla p+\nabla\cdot\boldsymbol{\tau}+\boldsymbol{F}_{\text{ext }}.\label{eq:mome}
\end{equation}
where $\mathrm{D}/\mathrm{D}t$ is the Lagrangian derivative, $\mathrm{D}/\mathrm{D}t\equiv\partial/\partial t+\boldsymbol{u}\cdot\nabla$,
$p$ is the pressure scalar field, $\boldsymbol{\tau}$ is the shear--stress
tensor, and $\boldsymbol{F}_{\text{ext }}$ is the vector field of
external forces that act on the fluid. Spatial and temporal dependency
for all terms in Eq. (\ref{eq:mome}) is implied. The shear--stress
tensor $\boldsymbol{\tau}$ is defined as:
\begin{equation}
\boldsymbol{\tau}=2\mu\boldsymbol{E},\label{eq:stress}
\end{equation}
where $\mu$ is the dynamic--viscosity scalar field, and the strain--rate
tensor $\boldsymbol{E}$ is defined as:
\begin{equation}
\boldsymbol{E}=\frac{1}{2}\left[\nabla\boldsymbol{u}+\left(\nabla\boldsymbol{u}\right)^{\top}\right],\label{eq:strain}
\end{equation}
where $^{\top}$ denotes the transpose operation. Within numerical
schemes, $\boldsymbol{u}$ for the next time step is obtained by solving
Eq. (\ref{eq:mome}), and directly discretising $\nabla\cdot\boldsymbol{\tau}$
yields inaccurate explicit schemes with rigorous time-step restrictions.
Therefore, the motivation for his work is establishing a strong form
of the momentum equation, that is convenient for direct numerical
manipulation and implicit solving.

\section{Divergence of shear stress}

The ``divergence of the scalar--tensor product'' identity for a
tensor $\mathbf{T}$ and scalar $\phi$ is defined as:
\begin{equation}
\nabla\cdot\left(\phi\mathbf{T}\right)=\phi\nabla\cdot\mathbf{T}+\mathbf{T}\,\nabla\phi.\label{eq:div-tensor}
\end{equation}
The divergence of shear--stress may be expanded by applying the identity
(\ref{eq:div-tensor}) to Eq. (\ref{eq:stress}):
\begin{align}
\nabla\cdot\boldsymbol{\tau} & =\nabla\cdot\left(2\mu\boldsymbol{E}\right)\nonumber \\
 & =2\mu\nabla\cdot\boldsymbol{E}+2\boldsymbol{E}\,\nabla\mu.\label{eq:divtau-start}
\end{align}
Eq. (\ref{eq:strain}) is substituted into Eq. (\ref{eq:divtau-start})
to obtain the following:
\begin{align}
\textrm{\ensuremath{\nabla\cdot\boldsymbol{\tau}}} & =\mu\,\nabla\cdot\left[\nabla\boldsymbol{u}+\left(\nabla\boldsymbol{u}\right)^{\top}\right]+2\boldsymbol{E}\,\nabla\mu\nonumber \\
\textrm{} & =\mu\nabla^{2}\boldsymbol{u}+\cancelto{\boldsymbol{0}}{\mu\,\nabla\cdot\left(\nabla\boldsymbol{u}\right)^{\top}}+2\boldsymbol{E}\,\nabla\mu\nonumber \\
\textrm{} & =\mu\nabla^{2}\boldsymbol{u}+2\boldsymbol{E}\,\nabla\mu.\label{eq:divtau-1}
\end{align}
\textbf{Remark 2.1.} The term $\nabla\cdot\left(\nabla\boldsymbol{u}\right)^{\top}$
is null vector, since the ``divergence of the transpose of a vector
gradient'' is equivalent to the ``gradient of the divergence of
a vector'', which is evident using the index notation:
\[
\nabla\cdot\left(\nabla\boldsymbol{u}\right)^{\top}=\frac{\partial^{2}u_{j}}{\partial x_{i}\partial x_{j}}=\frac{\partial}{\partial x_{i}}\frac{\partial u_{j}}{\partial x_{j}}=\nabla\left(\nabla\cdot\boldsymbol{u}\right)=\boldsymbol{0}.
\]

\section{Alternative expression for divergence of shear stress}

Alternatively, Eq. (\ref{eq:strain}) may be substituted into Eq.
(\ref{eq:stress}) and $\nabla\cdot\boldsymbol{\tau}$ may be expressed
as:
\begin{align}
\nabla\cdot\boldsymbol{\tau} & =\nabla\cdot\left(2\mu\boldsymbol{E}\right)\nonumber \\
 & =\nabla\cdot\left\{ \mu\left[\nabla\boldsymbol{u}+\left(\nabla\boldsymbol{u}\right)^{\top}\right]\right\} \nonumber \\
 & =\nabla\cdot\left(\mu\nabla\boldsymbol{u}\right)+\nabla\cdot\left[\mu\left(\nabla\boldsymbol{u}\right)^{\top}\right].\label{eq:divtau-start-alt}
\end{align}
The identity (\ref{eq:div-tensor}) may be applied to the second term
of the right-hand-side in Eq. (\ref{eq:divtau-start-alt}), therefore
obtaining:
\begin{align}
\nabla\cdot\boldsymbol{\tau} & =\nabla\cdot\left(\mu\nabla\boldsymbol{u}\right)+\cancelto{\boldsymbol{0}}{\mu\,\nabla\cdot\left(\nabla\boldsymbol{u}\right)^{\top}}+\left(\nabla\boldsymbol{u}\right)^{\top}\nabla\mu.\label{eq:divtau-alt-expanded}
\end{align}
\textbf{Example 3.1.} The variable--coefficient Laplacian discretised
using second--order finite differences on a uniform $n$-dimensional
grid is defined using the summation form as:
\[
\nabla\cdot\left(c_{i}\nabla\phi_{i}\right)\coloneqq\sum_{j}\frac{c_{j}+c_{i}}{2}\frac{\phi_{j}-\phi_{i}}{h^{2}},
\]
where $j$ are stencil nodes--neighbours to the central node $i$,
and $h$ is the grid spacing. 

\section{Momentum equation formulations}

Due to two different ways of expressing $\nabla\cdot\boldsymbol{\tau}$
using Eqs. (\ref{eq:divtau-1}) and (\ref{eq:divtau-alt-expanded}),
the momentum Eq. (\ref{eq:mome}) can be written using the \emph{standard
Laplacian (with the strain-rate)} notation:

\begin{equation}
\frac{\mathrm{D}\left(\rho\boldsymbol{u}\right)}{\mathrm{D}t}-\mu\nabla^{2}\boldsymbol{u}=\left[\nabla\boldsymbol{u}+\left(\nabla\boldsymbol{u}\right)^{\top}\right]\nabla\mu-\nabla p+\boldsymbol{F}_{\text{ext }},\label{eq:mom-1}
\end{equation}
or using the novel \emph{variable--coefficient Laplacian (with a
part of the strain-rate)} notation:
\begin{equation}
\frac{\mathrm{D}\left(\rho\boldsymbol{u}\right)}{\mathrm{D}t}-\nabla\cdot\left(\mu\nabla\boldsymbol{u}\right)=\left(\nabla\boldsymbol{u}\right)^{\top}\nabla\mu-\nabla p+\boldsymbol{F}_{\text{ext }}.\label{eq:mom-2}
\end{equation}
Completeness of including velocity terms on the left-hand-side defines
the accuracy and performance of a computational method. In order to
have flexible simulation schemes, $\mu$ and $\nabla\mu$ are computed
before solving Eq. (\ref{eq:divtau-1}) for $\boldsymbol{u}$ \citep{Pacheco2021,Basic2021}.
Due to this aspect and complexity of discretising $\boldsymbol{E}\nabla\mu$
product implicitly, it is commonly imposed on the right-hand-side.
This limits the time-step. However, Eq. (\ref{eq:divtau-alt-expanded})
allows to transfer a part of the strain rate information on the left-hand-side
without adding numerical complexity compared to Eq. (\ref{eq:divtau-1}).
Eq. (\ref{eq:mom-2}) is a convenient representation for computational
methods that discretise the strong formulation (e.g. finite difference
methods), because the variable--coefficient Laplacian and standard
Laplacian usually depend on the same discretisation procedure \citep{Basic2018jcp,Tzou2019,Gibou2002b}.
In the discrete context, the continuity Eq. (\ref{eq:continuity})
cannot be truly satisfied; in the best case it is satisfied to the
machine precision. The commonly used momentum Eq. (\ref{eq:mom-1})
that is derived based on Eq. (\ref{eq:divtau-1}) neglects the velocity
divergence due to Eq. (\ref{eq:continuity}), i.e. retains the complete
strain rate information on the right-hand-side. On the contrary, the
term $\nabla\cdot\left(\mu\nabla\boldsymbol{u}\right)$ in Eq. (\ref{eq:divtau-alt-expanded}),
which is used to derive the new momentum Eq. (\ref{eq:mom-2}), implicitly
includes part of the strain-rate information on the left-hand-side,
therefore having better numerical balance between left-hand-side and
right-hand-side.

\section{Concluding remarks}

This paper presents a derivation of two expressions for the momentum
equation of the generalised Navier-Stokes equation in strong form.
Due to the computational complexity of $\boldsymbol{E}\nabla\mu$
in the so-called Laplacian formulation, it is often translated to
the right-hand-side when solving the momentum equation for the velocity,
therefore sacrificing accuracy and limiting time step sizes. The newly
derived formulation of the momentum equation allows to use part of
the strain information using the same Laplacian numerics on the left-hand-side,
therefore potentially making future computational schemes simpler
with larger time-step sizes, while retaining the solving performance.

\bibliographystyle{unsrt}
\bibliography{0_home_josip_Documents_bibtex_Momentum}

\end{document}